\documentclass[aps,pre,amsmath,amssymb,superscriptaddress,floatfix,showpacs]{revtex4}
\newcommand{\beq}{\begin{eqnarray}}
\newcommand{\eeq}{\end{eqnarray}}
\renewcommand{\vec}[1]{{\mathbf{#1}}}

\newcommand{\react}[2]{\stackrel{#1}{\rightarrow}#2}
\newcommand{\der}[2]{\frac{\partial #1}{\partial #2}}

\newcommand{\dint}[3]{\int_#1^#2{\mathrm d}#3}

\begin{document}
\title{Aging processes in reversible reaction-diffusion systems}

\author{Vlad Elgart}
\affiliation{Department of Physics, Virginia Polytechnic Institute and State University,
 Blacksburg, Virginia 24061-0435, USA}
\author{Michel Pleimling}
\affiliation{Department of Physics, Virginia Polytechnic Institute and State University,
 Blacksburg, Virginia 24061-0435, USA}

\date{\today}

\begin{abstract}
Reversible reaction-diffusion systems display anomalous dynamics characterized by a power-law relaxation toward
stationarity. 
In this paper we study in the aging regime the nonequilibrium dynamical properties of some model systems with
reversible reactions.
Starting from the exact Langevin equations describing these models, we derive expressions for two-time
correlation and autoresponse functions and obtain a simple aging behavior for these quantities.
The autoresponse function is thereby found to depend on the specific nature of the chosen perturbation
of the system.

\end{abstract}


\maketitle

\section{Introduction}
The intriguing aging processes observed in nonequilibrium systems with slow (i.e. nonexponential) dynamics
have been the focus of many intensive research efforts in the past. Whereas the initial studies almost
exclusively focused on systems like glasses \cite{Stru78} and spin glasses \cite{Vin07},
it has been realized
quite recently that aging processes do not uniquely characterize these complex systems, but that they are also
encountered in much simpler situations \cite{Hen07}. Thus, the study of aging phenomena has for example been extended toward
magnetic systems \cite{Cug02,Cal05,Hen07a}, prepared initially in a disordered high temperature state and
then quenched to or below
their equilibrium critical point, and toward reaction-diffusion systems quenched to their
nonequilibrium critical point \cite{Hen07b}.
These investigations of simple systems displaying aging have led to an increased theoretical understanding of aging
processes taking place far from equilibrium.

Diffusion-limited irreversible reactions are often characterized by the presence of an absorbing phase transition,
separating an active phase from an inactive or absorbing state from which the system can not escape
\cite{Hin00,Odo04,Hen08}. These nonequilibrium phase transitions have attracted much interest, and different universality
classes have been identified \cite{Odo04,Elg06}. Various quantities, as for example the density of particles, display
simple power-laws when approaching these nonequilibrium critical points. In addition, the dynamical correlation length
increases as a power-law of time, similar to what is observed at an equilibrium critical point, revealing the
presence of slow dynamics. Consequently, aging processes have been studied in irreversible reaction-diffusion
systems quenched to their nonequilibrium critical point \cite{Ram04,Ens04,Bau05,Hin06,Odo06,Bau07,May07}.
Interestingly, two-time quantities, like the autocorrelation function $C(t,s)$
and the autoresponse function $R(t,s)$, display in these systems the same simple scaling behavior as the corresponding quantities
in equilibrium critical systems:
\begin{eqnarray}
C(t,s) & = & s^{-b} f_C(t/s) \label{autocorr_scal} \\
R(t,s) & = & s^{-1-a} f_R(t/s) \theta(t-s) \label{autoresp_scal}
\end{eqnarray}
where $a$ and $b$ are nonequilibrium exponents, whereas the scaling functions $f_C$ and $f_R$ only depend on the ratio
$t/s$, with $f_C(y) \sim y^{-\lambda_C/z}$ and $f_R(y) \sim y^{-\lambda_R/z}$ for $y \gg 1$. Here $z$ is the dynamical exponent,
and $\lambda_C$ and $\lambda_R$ are called autocorrelation and autoresponse exponents.
Finally, the step function $\theta(t-s)$ in the expression for the autoresponse function ensures causality.
The absence of detailed balance in irreversible reaction-diffusion systems
reveals itself mainly by the fact that the relation $a = b$, trivially observed at
equilibrium critical points, is no longer fulfilled (see \cite{Hen07b} for a recent review containing a thorough
discussion of this point).

All these recent studies investigated aging at nonequilibrium phase transitions.
This is of course due to the fact that in reaction-diffusion systems with irreversible reactions nonexponential
relaxation is exclusively encountered at absorbing phase transitions.

It is very remarkable that non-exponential
relaxation is {\it generically} found in {\it reversible} reactions, and this without fine-tuning
of the system parameters (as would be needed in order to be exactly at a phase transition point).
A power law behavior in the long time limit was
first predicted in \cite{Zel77} for the bimolecular reversible reaction
$A+B \rightleftharpoons C$ taking place in solutions, based on
physical arguments involving spatial concentration fluctuations.
This power law approach to stationarity in reversible reaction-diffusion systems
was later verified through more elaborated approaches \cite{Zel77b,Kan85,Osh89,Osh89b},
through numerical simulations \cite{Agm94,Agm95}, and through
some exactly solved models \cite{Rey99,Gop00,Agm00,Gop02}.
In addition, this power law behavior has been observed experimentally in
excited-state proton transfer reactions \cite{Hup92,Sol01,Pin01}.
All these studies demonstrate that the most important ingredient for aging, namely
slow dynamics, is typically encountered in reversible reaction-diffusion systems.

We propose to extend the study of aging phenomena to these reversible reaction-diffusion models.
Starting from idealized reaction-diffusion models, we exploit the fact that a set of
Langevin equations, describing the time evolution of the system, can be derived exactly for these systems \cite{Rey99}.
Within the standard field theoretical
representation of reaction-diffusion models we derive exact expressions for
the two-time correlation and autoresponse functions in the aging regime. For all the studied
models we recover a simple aging behavior where the scaling functions of these two-time quantities only depend
on the ratio of the two times. The most remarkable result obtained in this study concerns the
autoresponse function whose expression is found to depend on the nature of the perturbation applied to the system.

The paper is organized as follows. In Section \ref{section_2A-C}, we consider the reaction $A+A\rightleftharpoons C$.
Starting from the master equation description of the model, we use the exact map to a set of two Langevin equations
for some random complex variables $a$ and $c$. The concentrations $n_a(t)$ and $n_c(t)$ of the particles of types $A$ and $C$
are then given by the average of the random variables $a$ and $c$ over the complex noise: $n_a=\langle a \rangle$
and $n_c=\langle c \rangle$. Exploiting the controlled approximation scheme established in \cite{Rey99},
we derive exact expressions for two-time correlation and autoresponse functions in the dynamical scaling or aging regime.
In Section \ref{section_other}, we extend our study to the reactions $A+B\rightleftharpoons C$ and
$A+B\rightleftharpoons C + D$. Using the same approach as in Section \ref{section_2A-C}, 
we find the functional dependence
of the two-time quantities also for these reactions. In Section \ref{conclusions} we summarize our findings and discuss
open problems. Two additional technical points are discussed in the Appendices.

\section{The $A+A \rightleftharpoons C$ reaction scheme}\label{section_2A-C}

\subsection{Model and exact Langevin equations}

Following \cite{Rey99},
we consider two types of particles (called $A$ and $C$ particles) that diffuse on a $d$-dimensional hypercubic lattice.
Allowing multiple occupancy of a lattice site, particles at the same site may undergo the following reactions:
(1) the reaction $A + A \react{\lambda_0}{C}$ with rate $\lambda_0$ where two $A$ particles recombine to form a $C$ particle
and (2) the reverse reaction $C\react{\mu}{A+A}$ where a $C$ particle dissociates with rate $\mu$. These microscopic
rules are readily translated into the following master equation for the probability $P\left( \{m\},\{n\};t \right)$
(where $\{m\} = \{ \cdots, m_i, \cdots \}$ resp. $\{n\} = \{ \cdots, n_i , \cdots \}$ are the occupation numbers of
particles $A$ resp. $C$ for every lattice site $i$) that the configuration $\{m\},\{n\}$ is found at time $t$:
\begin{equation}\label{ME}
    \der{}{t}P\left( \{m\},\{n\};t \right) = \sum\limits_{i} {\cal H}_i P\left( \{m\},\{n\};t \right)
+ \sum\limits_{i} \sum\limits_{j(i)} D_{i,j} P\left( \{m\},\{n\};t \right)~.
\end{equation}
Here the term ${\cal H}_i P\left( \{m\},\{n\};t \right)$ contains the reactions taking place
at lattice site $i$ and is given by
\begin{eqnarray} \label{ME_singlepart}
    {\cal H}_i P(m_i,n_i;t) & = & \mu \left[(n_i+1)P(m_i-2,n_i+1;t) - n_i P(m_i,n_i;t)\right]\nonumber\\
   &&  + \lambda_0 \left[(m_i+2)(m_i+1)P(m_i+2,n_i-1;t) - m_i(m_i-1)P(m_i,n_i;t)\right]
\end{eqnarray}
where for convenience we wrote as arguments of $P$ only the occupation numbers $m_i$ and $n_i$ of lattice site $i$
that are changed by these reactions.
Diffusion processes are captured by the second term in Eq. (\ref{ME}) where the sum over $j(i)$ is a sum over the nearest
neighbor sites $j$ of the lattice site $i$. Indicating again only the occupation numbers that are changed in the process,
we have the following expression for $D_{i,j} P$:
\begin{eqnarray} \label{diff}
    {\cal D}_{i,j} P & = & \frac{D_a}{\ell^d} \left[(m_j+1)P(m_i-1,m_j+1;t) -  m_iP(m_i,m_j;t) \right]\nonumber \\
&& + \frac{D_c}{\ell^d} \left [(n_j+1)P(n_i-1,n_j+1;t) -  n_iP(n_i,n_j;t)\right]
\end{eqnarray}
where $D_a$ and $D_c$ are the diffusion constants of the $A$ and $C$ particles and $\ell$ is the lattice constant.

The master equation (\ref{ME}) must be supplemented by initial conditions. We here consider the case of an uncorrelated Poisson
distribution on each site and for each particle species. Prepared in this initial state, 
the system evolves toward chemical equilibrium
in the long time limit \cite{Rey99}. 


As shown by Rey and Cardy \cite{Rey99} the dynamics of this model allows an exact description in terms of a set of coupled
stochastic Langevin equations. Introducing $\lambda = \lambda_0 \ell^d$, one gets
\begin{eqnarray}
    (\partial_t - D_a\nabla^2) a({\bf x},t)
    & = & -2\lambda a^2({\bf x},t) + 2\mu c({\bf x},t) + \zeta({\bf x},t)
    \label{ALan} \\
    (\partial_t - D_c\nabla^2) c({\bf x},t)
    & = & \lambda a^2({\bf x},t) - \mu c({\bf x},t), \label{CLan}
\end{eqnarray}
where $\zeta$ is a \emph{complex} Gaussian noise with zero mean value whose
correlation is given by
\begin{equation}\label{noise}
    \langle \zeta({\bf x},t)\zeta({\bf x}',t') \rangle
    = 2\langle \mu c({\bf x},t) - \lambda a^2({\bf x},t) \rangle
    \delta({\bf x}-{\bf x}') \delta(t-t').
\end{equation}
Here, the vector ${\bf x}$ describes the $d$-dimensional space coordinates, whereas the bracket notation stands
for the average over the noise. Inserting Eq. (\ref{CLan}) into this expression yields
\begin{equation}\label{noise_corr}
    \langle \zeta({\bf x},t)\zeta({\bf x}',t') \rangle
    = -2\partial_t\langle c(t)\rangle\delta({\bf x}-{\bf x}') \delta(t-t')
\end{equation}
for spatial homogeneous initial conditions.

Note that the variables $a({\bf x},t)$ and $c({\bf x},t)$ do not represent the particle densities, as they are
complex \cite{Rey99}. The mean densities $n_a(t)$ and $n_c(t)$ of the particles of types $A$ and $C$, which are
of course real valued, are given by the averages $\langle a({\bf x},t) \rangle$ and $\langle c({\bf x},t) \rangle$
of these complex variables over the noise.

As the density $n_c(t) = \langle c({\bf x},t) \rangle$ reaches a stationary value in the limit $t \longrightarrow \infty$,
the noise correlation, see Eq. (\ref{noise_corr}), will vanish in the long time limit. As a consequence of the vanishing
of the fluctuations at equilibrium one can compute the actual values of the equilibrium densities, which are given by
their mean field values. These equilibrium densities $a_\infty$ and $c_\infty$ satisfy the relationship
\begin{equation}
    \lambda a_\infty^2 = \mu c_\infty\,, \label{steady}
\end{equation}
which follows directly from Eq. (\ref{noise}).

It is important to notice that the dynamics conserves the quantity $n_a(t)+ 2 n_c(t)$ due to the total mass conservation. In the case of equal diffusion constants $D_a=D_c\equiv D$, which is the case we discuss in the following,
we have in addition that $\chi = a + 2c$ obeys the noisy diffusion equation
\begin{equation}\label{cons}
    (\partial_t - D\nabla^2) \chi({\bf x},t) = \zeta({\bf x},t)~,
\end{equation}
as follows directly from Eqs. (\ref{ALan}) and (\ref{CLan}). Since $\langle\zeta\rangle=0$ and the
initial conditions are homogeneous, the average value of the field $\langle\chi({\bf x},t)\rangle$ is conserved.
In particular we have
$\chi_0\equiv \langle\chi({\bf x},0)\rangle = \langle\chi({\bf x},\infty)\rangle\equiv \chi_\infty$.
Using Eq. (\ref{steady}) one can therefore express the equilibrium densities as a function
of $\chi_0$ \cite{Rey99}:
\begin{eqnarray}\label{params}
    a_\infty = \frac{\mu}{4\lambda}\left(\sqrt{1+\frac{8\lambda}{\mu} \chi_0} - 1\right)~~~\mbox{and}~~~
    c_\infty = \frac{1}{2}(\chi_0 - a_\infty)\,.
\end{eqnarray}

Starting from this Langevin description and exploiting the existence of the conserved quantity, Rey and Cardy \cite{Rey99}
developed a systematic approximation scheme that enabled them not only to derive the power-law relaxation toward
equilibrium but also to compute the corresponding amplitude exactly. In the following we extend this approach to
two-time quantities and derive in leading order exact expressions for correlation and response functions in the aging regime.

\subsection{Two-time correlation function}
New insights into the behavior far from stationarity can be obtained through the analysis of the connected correlation
function
\begin{equation} \label{C}
C(t,\vec{x};s,\vec{y}) = \langle \delta c(\vec{x},t) \, \delta c(\vec{y},s) \rangle - \langle \delta c(\vec{x},t) \rangle
\, \langle \delta c(\vec{y},s) \rangle~,
\end{equation}
with $\delta c(\vec{x},t) = c(\vec{x},t) - \langle c(\vec{x},\infty) \rangle = c(\vec{x},t) - c_\infty$. In the definition
of the correlation we have taken into account that in the stationary state the mean value of the variable $c$ is
$c_\infty$. In case $\vec{x} = \vec{y}$, we are dealing with the autocorrelation function $C(t,s) =
C(t,\vec{x};s,\vec{x}) = C(t,\vec{0};s,\vec{0})$, where in the last identity we exploited the spatial
homogeneity of our system.

In principle, we should define a similar quantity for the variable $a(\vec{x},t)$. However, as
$\delta a(\vec{x},t) = a(\vec{x},t) - a_\infty = \delta \chi(\vec{x},t) - 2 \, \delta c(\vec{x},t)$ with
$\delta \chi(\vec{x},t) = \chi(\vec{x},t) - \chi_\infty$, we can immediately derive the correlator
for $a$ once we know the correlators for the conserved quantity $\chi$ and for the variable $c$.

Starting point for the computation of the correlation function is the following Langevin equation
for the quantity $\delta c(\vec{x},t)$ \cite{Rey99}:
\begin{eqnarray}
    (\partial_t - D\nabla^2 + \sigma)\,\delta c(\vec{x},t) & = & 4\lambda\,\delta c^2(\vec{x},t) -
4\lambda\,\delta \chi(\vec{x},t) \,\delta c(\vec{x},t) + \lambda\,\delta \chi^2(\vec{x},t) \nonumber \\
&& + \frac{1}{2} (\sigma-\mu)\,\delta \chi(\vec{x},t)\,, \label{deltac}
\end{eqnarray}
where $\sigma\equiv 4\lambda a_\infty + \mu$.

The formal solution of this equation is
\begin{equation}\label{formsol}
    \delta c = G[(c-c_0)\delta(t) + 4\lambda\,\delta c^2 - 4\lambda\,\delta \chi\,\delta c + \lambda\,\delta \chi^2
+ (\sigma-\mu)\,\delta \chi]\,,
\end{equation}
with the Green function
\begin{equation}\label{defG}
    G[f](\vec{x},t) = \dint{0}{t}{t'}\! \int\mathrm{d}^d x'\, e^{-\sigma(t-t')} G_0(\vec{x}-\vec{x}',t-t') f(\vec{x}',t')\,.
\end{equation}

The nonlinear equation (\ref{formsol}) is solved by the systematic approximation scheme developed in \cite{Rey99}.
As a result we obtain to leading order the following expressions for $\langle\delta c(\vec{x},t) \rangle$ and
$\langle\delta c(\vec{x},t) \delta c(\vec{y},s) \rangle$:
\begin{eqnarray}
    \langle\delta c(\vec{x},t)\rangle & = & \frac{\lambda\mu^2}{\sigma^3} \langle\delta
\chi^2(\vec{x},t)\rangle\,,\label{density}\\
\langle\delta c(\vec{x},t) \, \delta c(\vec{y},s)\rangle & = & \left(\frac{\sigma-\mu}{2\sigma} \right)^2\langle\delta
\chi(\vec{x},t) \, \delta \chi(\vec{y},s)\rangle\,.\label{var}
\end{eqnarray}
It follows that the calculation of our correlation function reduces to the calculation of the expectation values
$\langle \delta \chi^2(\vec{x},t) \rangle$ and $\langle \delta \chi(\vec{x},t) \, \delta \chi(\vec{y},s)
\rangle$. Rey and Cardy already computed the
first quantity and obtained in leading order
\begin{equation} \label{chi2}
\langle \delta \chi^2(\vec{x},t) \rangle = 2 (c_0 - c_\infty ) \left( 8 \pi D t \right)^{-d/2}~.
\end{equation}
For the computation of $\langle \delta \chi(\vec{x},t) \, \delta \chi(\vec{y},s) \rangle$ we exploit the
fact that $\chi(\vec{x},t)$
obeys the noisy diffusion equation (\ref{cons}) with the noise correlator given by Eq. (\ref{noise_corr}). This readily
yields the expression
\begin{equation}\label{chi}
    \langle\delta \chi(\vec{x},t)\, \delta \chi(\vec{y},s)\rangle = -2\dint{0}{t}{t_1}\int\mathrm{d}^d x_1
G_0(\vec{x}-\vec{x}_1,t-t_1)G_0(\vec{y} -\vec{x}_1,s-t_1)\partial_t \langle c(t_1)\rangle\,,
\end{equation}
where $G_0({\bf x},t)$ is the free propagator:
\begin{equation}
    G_0({\bf x},t) = \theta(t)\,(4\pi Dt)^{-d/2}\exp\left(-\frac{x^2}{4Dt}\right)\,,
\end{equation}
and $\theta(t)$ is the step function.
Hence
\begin{equation}\label{conserved_corr}
    \langle\delta \chi(\vec{x},t)\, \delta \chi(\vec{y},s)\rangle = -2\dint{0}{s}{t_1}\left[4\pi D(t+s-2t_1)\right]^{-d/2}
\exp\left\{-\frac{(\vec{x}-\vec{y})^2}{4D(t+s-2t_1)} \right\}\partial_t \langle c(t_1)\rangle\,.
\end{equation}
For $\vec x = \vec y$ we obtain up to some numerical prefactor an integral that Rey and Cardy already discussed
in \cite{Rey99}. 
In the aging regime, where both $t$ and $s$ are large, this then leads to the following leading behavior:
\begin{equation}
    \langle\delta \chi(\vec x,t)\delta \chi(\vec x,s)\rangle =
    2(c_0 - c_\infty)\left[4\pi D(t+s)\right]^{-d/2} ~.
\end{equation}
We therefore have for the autocorrelation function in the aging regime the expression
\begin{eqnarray}
    C(t,s) & = & \langle \delta c(\vec{x},t) \, \delta c(\vec{x},s) \rangle - \langle \delta c(\vec{x},t) \rangle
\, \langle \delta c(\vec{x},s) \rangle \nonumber \\
& = & \left(\frac{\sigma-\mu}{2\sigma} \right)^2\langle\delta
\chi(\vec{x},t) \, \delta \chi(\vec{y},s)\rangle - \left( \frac{\lambda\mu^2}{\sigma^3} \right)^2 \langle\delta
\chi^2(\vec{x},t)\rangle \langle\delta \chi^2(\vec{x},s)\rangle \nonumber \\
& = & \frac{1}{2} \left(\frac{\sigma-\mu}{\sigma} \right)^2 (c_0 - c_\infty)\left[4\pi D(t+s)\right]^{-d/2} \label{corr_func}~,
\end{eqnarray}
where in the last line we omitted all sub-leading correction terms. It is interesting to notice that the term
$\langle \delta c(\vec{x},t) \rangle \, \langle \delta c(\vec{x},s) \rangle$ is of the order $s^{-d/2}t^{-d/2}$ and
therefore only contributes to the correction terms. This reflects the fact (already noticed in \cite{Rey99})
that the quantity $c$ is not a self-averaging quantity.

The derived expression for the autocorrelation function can be cast in the form $C(t,s) = s^{-b} f_C(t/s)$ that characterizes
a simple aging behavior. This also yields the following values for the nonequilibrium exponents (see Eq. (\ref{autocorr_scal})):
\begin{equation}
b= d/2 ~~~\mbox{and} ~~~ \lambda_C/z = d/2~.
\end{equation}

The space-time correlation function $C(t,\vec{x};s,\vec{y})$ can
be computed in exactly the same way, with the final result
\begin{equation}
   C(t,\vec{x};s,\vec{y}) = \frac{1}{2}\left(\frac{\sigma-\mu}{\sigma}\right)^2
   (c_0 - c_\infty)\left[4\pi D(t+s)\right]^{-d/2}\exp\left\{-\frac{(x-y)^2}{4D(t-s)} \right\}~.
\end{equation}

\subsection{Two-time response functions}\label{section_response}

The system can be perturbed in several possible ways in order to compute the linear response. One of the possibilities
is to inject new particles (which can be particles of either type $A$ or $C$) at time $s$. The `injection'
process is assumed to be random with the same small occurrence probability at each lattice site.
The response of the system to that perturbation is then monitored at a later time $t$
by measuring the densities of particles of type $A$ or $C$. In this way we obtain different
responses that we note as $R_i^f(t,s)$ where $i$ stands for the type of particles that are created whereas
$f$ indicates the type of particles whose density is measured. Thus, the response $R_C^A(t,s)$, formally given
by the equation
\begin{equation}
R_C^A(t,s) = \left. \frac{\delta \langle a \rangle(t)}{\delta h_C(s)} \right|_{h_C \longrightarrow 0}~,
\end{equation}
means that we are measuring
the linear response of the $A$ particles density to the additional creation of $C$ particles only. It is important
to notice that this process violates the conservation of the quantity $\langle \chi \rangle$.
In order to assess the impact of this violation on the response of the system, we also
consider a process that conserves the total mass of the particles. As we discuss in the following, the
response of a reversible diffusion-reaction system strongly depends on the chosen perturbation.

Let us start by injecting particles of type $C$ into the system and by monitoring the subsequent change in the particle density
of the same particle type.
A creation process $\emptyset\react{h_C}{C}$ modifies the single site part of the master equation (\ref{ME}) which now reads as
\begin{equation}\label{ME_response_0d}
    {\cal H}_i^h P(m_i,n_i;t) = {\cal H}_iP(m_i,n_i;t) + h_C(t)\left[P(m_i,n_i-1;t) - P(m_i,n_i;t)\right]\,,
\end{equation}
where ${\cal H}_iP(m_i,n_i;t)$ is the expression (\ref{ME_singlepart}) one has without additional creation processes.

We might want to inject $\Omega_C$ additional $C$ particles at time $s$, so that
the particle injection probability $h_C(t)$ is given by
\begin{equation}\label{h0}
h_C(t) = \Omega_C \delta (t-s)~.
\end{equation}
However, as the final state of the evolution will again be a homogeneous state, the concrete form of $h_C$ is not
of real importance as long as the particle injection has ended before the measurement of the response at 
time $t$, with the total number of injected particles being given by
\begin{equation}\label{omega}
    \Omega_C = \dint{0}{t}{\tau}\,h_C(\tau)~.
\end{equation}


With this particle injection process the Langevin equation for the $C$ particles is now given by
\begin{equation}\label{CLan_response}
    (\partial_t - D\nabla^2) c^h(\vec{x},t) =
    \lambda a^h(\vec{x},t)^2 - \mu c^h(\vec{x},t) + h(t)\,,
\end{equation}
where in the continuum limit we set
\begin{equation}\label{h}
    h(t) = \frac{h_C(t)}{\ell^d}~~~ \mbox{and}~~~ \Omega = \frac{\Omega_C}{\ell^d}\,,
\end{equation}
thereby dropping the index $C$ for the quantities divided by the volume of the system.
We use in the following the index $h$ in order to emphasize the presence of the additional creation process and to distinguish
the corresponding quantities from those obtained in the absence of this process.
The Langevin equation for the $A$ particles is unchanged by the creation of $C$ particles,
but the noise-noise correlator (\ref{noise_corr}) becomes
\begin{equation}\label{noise_corr_response}
    \langle \zeta(\vec{x},t)\zeta(\vec{x}',t') \rangle = 2\left[h(t) -
    \partial_t\langle c^h(t)\rangle\right]\delta(\vec{x}-\vec{x}') \delta(t-t')~,
\end{equation}
for spatial homogeneous initial conditions,
due to the Eqs. (\ref{noise}) and (\ref{CLan_response}).

With this one readily verifies the validity of the following set of algebraic equations:
\begin{eqnarray}
    a_\infty^h + 2 c_\infty^h & = & a_0 + 2 c_0 + 2\Omega = \chi_0 + 2\Omega~,\label{mass}\\
    \lambda \left(a_\infty^h\right)^2 & = & \mu c_\infty^h~.\label{eq}
\end{eqnarray}
Note that as $\Omega$ has the meaning of the average total change in the number of $C$ particles due to the creation process,
Eq. (\ref{mass}) reflects the modification of the total mass due to this process. Equations (\ref{mass}) and
(\ref{eq}) immediately
yield the following expressions for the mean values $a_\infty^h$ and $c_\infty^h$ in the new stationary state:
\begin{eqnarray}\label{paramsh}
    a_\infty^h = \frac{\mu}{4\lambda}\left(\sqrt{1+\frac{8\lambda}{\mu} (\chi_0 + 2 \Omega)} - 1\right)~~~\mbox{and}~~~
    c_\infty^h = \frac{1}{2}(\chi_0 +2 \Omega - a_\infty)\,
\end{eqnarray}
which differ from the expressions (\ref{params}) without the field $h$ through the replacement of $\chi_0$ by $\chi_0 + 2
\Omega$.

As the coefficients $a_\infty^h$ and $c_\infty^h$ are
again constants,
it follows that also in this case the perturbation series derived by Rey and Cardy \cite{Rey99} and sketched
in the previous Section only involves time independent coefficients. Therefore, any quantity
which is a function of the coefficients $a_\infty^h$ and $c_\infty^h$ can be derived in exactly the same manner
as discussed before. 
For instance, the constant $\sigma$ becomes
\begin{eqnarray}\label{AA_corr}
    \sigma^h = 4 \lambda a_\infty^h + \mu~.
\end{eqnarray}
Hence, in order to calculate the $h$-dependent density $\langle c^h\rangle \equiv c_\infty^h + \langle\delta\!c^h\rangle$
we just need to substitute into Eq. (\ref{density}) the expression (\ref{AA_corr}) and the $h$-dependent
expression for $\langle \delta \chi^h \rangle$.
This last quantity is obtained by noicing that due to the presence of the additional injection process
the Langevin equation for $\chi$ now reads
\begin{equation}\label{chi_modified}
    \left(\der{}{t} + \nabla^2\right)\chi^h(\vec{x},t)
     = \zeta(\vec{x},t) + 2h(t)\,,
\end{equation}
This yields for $\langle (\delta \chi^h)^2 \rangle$ the expression
\begin{equation}\label{chi_corr_modified}
    \langle(\delta \chi^h)^2\rangle = 2 ( c_0 - c^h_\infty ) (8\pi D t)^{-d/2} +
    2 \dint{0}{t}{\tau}\,\left[ 8\pi D(t-\tau) \right]^{-d/2}h(\tau)
\end{equation}
in leading order.

The last remaining step is to calculate the two-time linear response function
\begin{equation}
    R_C^C(t,s) =\left. \frac{\delta\langle c^h(t)\rangle}{\delta h(s)} \right|_{h \longrightarrow 0}\,,
\end{equation}
which after some elementary algebra is given by the expression
\begin{eqnarray}\label{response}
    R_C^C(t,s) & =  & 2 \frac{\lambda \mu^2}{\sigma^3} \left[ \frac{\mu}{\sigma}-1 -
\frac{24 \lambda \mu}{\sigma^2}
\left( c_0 - c_\infty \right) \right] (8\pi D t)^{-d/2} \theta(t-s) \nonumber\\
&& +  2 \frac{\lambda \mu^2}{\sigma^3} \left[8\pi D(t-s)\right]^{-d/2} \theta(t-s) \,,
\end{eqnarray}
where we used the fact that
$\frac{\delta \Omega}{\delta\,h(s)} = \theta(t-s)$. This expression can also be cast in the standard
scaling form $R_C^C(t,s) = s^{-1-a} f_R(t/s) \theta(t-s)$, with $a=d/2-1$ and $\lambda_R/z = d/2$. Interestingly,
we have $a \neq b$, even though the system evolves toward chemical equilibrium. We also note that $\lambda_R = \lambda_C$.

The change in the density of particles A due to the injection of $C$ particles, $R_C^A$, is related to the response
$R_C^C$ by
\begin{equation} \label{rca}
R_C^A(t,s) = 2 \theta(t-s) - 2 R_C^C(t,s)~,
\end{equation}
which follows directly from $\langle a^h \rangle + 2 \langle c^h \rangle = \langle \chi^h \rangle = \chi_0 + 2 \Omega$.
Thus the expression (\ref{rca}) is the sum of two terms with different scaling behaviors where the constant term is the
leading one.

Let us now discuss the response of the system to an injection of particles of type A. One possible way of doing this
consists in injecting pairs of $A$ particles into the system with a rate $h_{2A}$: $\emptyset\react{h_{2A}}{2A}$. As
these pairs can immediately react to form a $C$ particle, it is expected that this leads to the same behavior as observed
when injecting $C$ particles into the system. Indeed, the creation of pairs of $A$ particles on the one hand changes
the noise-noise correlator which now reads (with $h = h_{2A}/l^d$)
\begin{equation}\label{noise_corr_response2}
    \langle \zeta(\vec{x},t)\zeta(\vec{x}',t') \rangle = \left[h(t) - 2
    \partial_t\langle c^h(t)\rangle\right]\delta(\vec{x}-\vec{x}') \delta(t-t')~,
\end{equation}
and on the other hand modifies the Langevin equation for the $A$ particles:
\begin{equation}
    (\partial_t - D_a\nabla^2) a({\bf x},t) = -2\lambda a^2({\bf x},t) + 2\mu c({\bf x},t) + \zeta({\bf x},t) + 2h(t)~.
\end{equation}
Using in addition that the Langevin equation for the $C$ particles remains unchanged, it is then straightforward to
show that we have $R_{2A}^C = R_C^C$ and $R_{2A}^A = R_C^A$.

The situation changes if instead of injecting pairs of $A$ particles only single $A$ particles are created with rate
$h_A$. In that case the additional $A$ particles do not automatically lead to the formation of additional
$C$ particles, but instead a newly created $A$ particle must first diffuse through the system in order to encounter
another additional $A$ particle.
Formally the creation of single $A$ particles again shows up in the Langevin equation (\ref{ALan}) for the $A$ particles as
an additional field term $h(t)$, but the noise-noise correlator and, subsequently, the $\delta \chi$ correlator
are the same as for the $h=0$ case:
\begin{eqnarray}
    \langle \zeta(\vec{x},t)\zeta(\vec{x}',t') \rangle & = & -2
    \partial_t\langle c^h(t)\rangle\delta(\vec{x}-\vec{x}') \delta(t-t')~,\\
    \langle(\delta \chi^h)^2\rangle & = & 2 ( c_0 - c^h_\infty ) (8\pi D t)^{-d/2}\,.
\end{eqnarray}
This then readily yields the expression
\begin{equation}\label{responseA}
    R_A^C(t,s)=
\frac{ \lambda \mu^2}{\sigma^3} \left[ \frac{\mu}{\sigma}-1 -
\frac{24 \lambda \mu}{\sigma^2}
\left( c_0 - c_\infty \right) \right] (8\pi D t)^{-d/2} \theta(t-s) \,,
\end{equation}
for the response of the $C$ particles to the creation of single $A$ particles. It is important to note
that the expression (\ref{responseA}) does not depend (at least in leading order) on the excitation time $s$.
This is a direct consequence of the fact that the rate of encounter of two newly created $A$ particles does not
depend on the time $s$ at which the particles have been created.

All the perturbations considered so far have in common that the quantity $\langle \chi \rangle =
n_a + 2n_c$, that would be constant
without the perturbation, has a different value in the stationary state than in the initial state. In order to assess the
importance of this conserved quantity, we also looked at the response to a perturbation that keeps $\langle \chi \rangle$
unchanged. This can be achieved by the simultaneous creation of particles of one type and destruction of particles of
the other type. For example, we can create pairs of $A$ particles with rate $h_\chi$ and at the same time remove
$C$ particles with rate $h_\chi$.
In principle, the removal procedure is an ill defined process, since one may end up with negative number of particles on
a given site.
However, for large equilibrium concentrations $a_\infty$ and $c_\infty$, we expect the number of particles on each site of the
lattice to remain larger than zero following an infinitesimal excitation of the type just described.

It is important to note that the Langevin equation (\ref{cons}) for $\chi$ is unaffected by this perturbation, even so
additional field terms are entering into the Langevin equations for $a$ and $c$. Doing the calculations along the lines
just sketched for the other perturbations, one remarks that, due to the fact that the asumptotic values $c_\infty^h$
and $a_\infty^h$ are independent of the field $h = h_\chi/l^d$, the response functions become time translational
invariant. Thus for the response of the $C$ particles to this perturbation we get
\begin{equation} \label{res_chi1}
    R_\chi^C(t,s)=
2 \frac{\lambda \mu^2}{\sigma^3} \left[8\pi D(t-s)\right]^{-d/2} \theta(t-s)~,
\end{equation}
whereas for the response of the $A$ particles we obtain
\begin{equation} \label{res_chi2}
R_\chi^A(t,s) = -2 R_\chi^C(t,s) = - 4 \frac{\lambda \mu^2}{\sigma^3} \left[8\pi D(t-s)\right]^{-d/2} \theta(t-s) ~,
\end{equation}
as $\Omega = 0$.

It has been proposed \cite{Hen01,Hen02} that in aging systems space-time symmetries can be exploited in order
to derive exact expressions for two-time quantities (see \cite{Hen07a} and \cite{Hen07b} for recent reviews of
this approach in the context of magnetic and of reaction-diffusion systems, respectively). Thus for the
autoresponse function, this theoretical approach, called the theory of local scale invariance, yields the
following general expression:
\begin{equation} \label{lsi}
R(t,s) =r_0 s^{-a-1}
\left(\frac{t}{s}\right)^{1+a' -
\lambda_R/z} \left(\frac{t}{s} -1 \right)^{-1-a'} \, \theta(t-s)~,
\end{equation}
where the values of the exponents $a$, $a'$ and $\lambda_R$ are not fixed by the theory, whereas $r_0$ is a nonuniversal
numerical prefactor and $z$ is the dynamical exponent. In the reversible reaction-diffusion system considered here,
we have the interesting situation that
we can design different perturbations of the system and monitor the reaction of the system to these perturbations.
Comparing the theoretical expression (\ref{lsi}) with our exact results, we observe that the responses (\ref{res_chi1})
and (\ref{res_chi2})
to a perturbation that does not change the conserved quantity $\langle \chi \rangle$ can indeed be cast in the form (\ref{lsi}),
with $a = a' = d/2 -1$ and $\lambda_R/z = d/2$. On the other hand, however, perturbations
that change the value of the conserved quantity can not be cast in the form (\ref{lsi}), indicating
that one must be careful when applying space-time symmetries to perturbations that change 
quantities that are otherwise conserved by the dynamics of the system.



\section{Other reversible reaction schemes}\label{section_other}
\subsection{The $A+B \rightleftharpoons C$ reaction scheme}
The treatment of the bimolecular reaction scheme $A+B \rightleftharpoons C$ closely follows
that of the scheme $A+A \rightleftharpoons C$ discussed in the previous Section. The main
difference is the presence of two 'conservation' laws for the average densities \cite{Rey99}:
\begin{equation}
a_\infty + b_\infty + 2 c_\infty = a_0 + b_0 + 2 c_0 ~~~\mbox{and} ~~~ a_\infty - b_\infty = a_0 - b_0~,
\end{equation}
where $a_0$, $b_0$, $c_0$, respectively $a_\infty$, $b_\infty$, $c_\infty$, are the concentrations
of the $A$, $B$, $C$ particles in the initial, respectively final, state. Writing down the exact
Langevin equations, one notices that the Langevin equation for $c(\vec{x},t)$ is still noise independent.
Exploiting this property, we find for the two-point correlation function the expression
\begin{equation}\label{corr_func2}
    C(t,s) = \frac{1}{2} \left( \frac{\sigma - \mu}{\sigma} \right)^2(c_0-c_\infty)(4\pi D(t+s))^{-d/2}\,,
\end{equation}
which is of the same form as for the $A+A \rightleftharpoons C$ reaction scheme, see Eq. (\ref{corr_func}).
The only difference is that the constant $\sigma$ is now given by $\sigma = \lambda ( a_0 + b_0 + 2 c_0 - 2
c_\infty) + \mu$. We also notice that the expressions for the different response functions are unchanged,
provided that the new expression for $\sigma$ is used.


\subsection{The $A+B \rightleftharpoons C+D$ reaction scheme}
More interesting is the reversible reaction $A+B \rightleftharpoons C+D$, not discussed in \cite{Rey99},
as here the Langevin equation for each reactant
does depend on noise. It is worth mentioning that the $A+B \rightleftharpoons C+D$ reaction scheme is readily found in
experimental situations, one well-known example being ethanoic acid dissolved in water that forms
ethanoate and hydronium ions: $CH_3CO_2H + H_2O \rightleftharpoons CH_3CO_2^- + H_3O^+$.
In order to make the following discussion more compact, we use the symbols $A_1$ and $A_2$ instead of $A$ and $B$, and
$C_1$ and $C_2$ instead of $C$ and $D$.


The exact Langevin equations for this four species reversible reaction read:
\begin{eqnarray}
    \left(\der{}{t} -D\nabla^2\right) a_i = f(a_i,c_i) + \zeta_{a_i}\,,\label{a}\\
    \left(\der{}{t} -D\nabla^2\right) c_i = -f(a_i,c_i) + \zeta_{c_i}\,,\label{c}
\end{eqnarray}
where $i=1,2$ and $f(a,c) \equiv (\mu c_1 c_2 - \lambda a_1 a_2)$. Here the complex variables are again related to the average
particle densities, e.g. $n_{a_1} = \langle a_1 \rangle$ is the mean density of the particles of type $A_1$.
The only non vanishing noise--noise correlators are
\begin{eqnarray}
    \langle\zeta_{a_1}({\bf r}) \zeta_{a_2}({\bf r}')\rangle = 2\delta({\bf r} - {\bf r}')
    \langle f(a,c)\rangle\,,\label{abcd_noise_a}\\
    \langle\zeta_{c_1}({\bf r}) \zeta_{c_2}({\bf r}')\rangle = -2\delta({\bf r} - {\bf r}')
    \langle f(a,c)\rangle\,.\label{abcd_noise_c}
\end{eqnarray}
There exist three `conserved' quantities, namely $\phi_t \equiv \sum_{i=1}^{2}(a_i + c_i)$, $\phi_a \equiv a_1-a_2$,
and $\phi_c \equiv c_1-c_2$,
which obey the noisy diffusion equations
\begin{equation}
    \left(\der{}{t} -D\nabla^2\right) \phi_j = \tilde\zeta_j
\end{equation}
with $j=t,a,b$. The noise terms $\tilde\zeta_j$ are thereby just linear combinations of the $\zeta_{a_i}, \zeta_{c_i}$:
$\tilde\zeta_t \equiv \sum_{i=1}^{2}(\zeta_{a_i} + \zeta_{c_i})$, $\tilde\zeta_a \equiv \zeta_{a_1}-\zeta_{a_2}$, and
$\tilde\zeta_c \equiv \zeta_{c_1}-\zeta_{c_2}$.

Let us now calculate the average density of, say, $A_1$ particles, $\langle a_1\rangle$. 
Since the solution of the steady state is given by the condition $f(a^\infty_i, c^\infty_i) = 0$, 
where $a^\infty_i$ and $c^\infty_i$ are the average particle densities in the steady state,
we can find all four coefficients $a^\infty_i$ and $c^\infty_i$ ($i=1,2$)
by exploiting the conservation laws $\langle \phi_t \rangle = \mathrm{const} \equiv \Phi_t$, 
$\langle \phi_a \rangle = \mathrm{const} \equiv \Phi_a$, and $\langle \phi_c \rangle = \mathrm{const} \equiv \Phi_c$:
\begin{eqnarray}
a^\infty_1 & = & \frac{\left(\Phi_t + \Phi_a\right)^2 - \Phi_c^2}{4\Phi_t} ~~~~,~~~~
a^\infty_2 = \frac{\left(\Phi_t + \Phi_c\right)^2 - \Phi_a^2}{4\Phi_t}\,, \nonumber \\
c^\infty_1 & = & \frac{\left(\Phi_t - \Phi_c\right)^2 - \Phi_a^2}{4\Phi_t} ~~~~,~~~~
c^\infty_2 = \frac{\left(\Phi_t - \Phi_a\right)^2 - \Phi_c^2}{4\Phi_t}\,.
\end{eqnarray}

We will in the following discuss to some extent the case $\mu = \lambda$ which has the virtue that the
algebra is quite easy. Moreover, by choosing an appropriate time rescale we can always set $\mu = 1$.
The general situation $\lambda \neq \mu$ can be treated along the same lines, but the algebra is rather
involved. As we find for the general case at late times the same functional dependence of the particle density 
as for the case
$\mu = \lambda$, we refrain from giving the details of this calculation and only quote the 
result at the end of this Section.

Substituting $a_i = a_i^\infty + \delta\! a_i$, $c_i = c_i^\infty + \delta\! c_i$, 
and $\phi_j = \Phi_j + \delta\! \phi_j$
(where $i=1,2$ and $j=t,a,c$) into Eq. (\ref{a}) yields, after some algebra, 
\begin{eqnarray} \label{abcd_a1}
    \left(\der{}{t} -D\nabla^2\right)\, \delta\! a_1 &=& \Phi_t\, \delta\! a_1 - \frac{\Phi_t^2 + \Phi_c^2 - \Phi_a^2}{4\Phi_t}\,\delta\! \phi_t - \frac{\Phi_t - \Phi_a}{2}\,\delta\!\phi_a + \frac{\Phi_c}{2}\,\delta\!\phi_c + \zeta_{a_1}\nonumber\\
    && - \frac{1}{4}(\phi_t^2 + \phi_a^2 - \phi_c^2) - \frac{1}{2}\delta\!\phi_t\,\delta\!\phi_a + \delta\! \phi_t\,\delta\! a_1\,.
\end{eqnarray}
Note that the above equation is linear in terms of the function $\delta\! a_1$. This is a consequence
of our choice $\mu = \lambda$, and is not true in the general case.

By introducing a new Green function
\begin{equation}
    G\equiv e^{-\Phi_t t}G_0(\vec{x},t)\,,
\end{equation}
we can set up a perturbation series along the same lines as discussed previously for the other reaction schemes. We 
then get
\begin{eqnarray}
    \langle\delta\! a_1\rangle & = & - G\left[\frac{1}{4}\langle\delta\! \phi_a^2 - \delta\! \phi_c^2\rangle + \langle\delta\! \phi_t^2\rangle - \frac{1}{2}\langle\delta\!\phi_t\,\delta\!\phi_a\rangle\right]\nonumber\\ 
&& +  G\left[\langle\phi_t\,G\left[- \frac{\Phi_t^2 + \Phi_c^2 - \Phi_a^2}{4\Phi_t}\,\delta\! \phi_t - \frac{\Phi_t - \Phi_a}{2}\,\delta\!\phi_a + \frac{\Phi_c}{2}\,\delta\!\phi_c + \zeta_{a_1}\right]\rangle\right] + \cdots\,.\label{abcd_ave}
\end{eqnarray}
where the dots refer to higher order terms that are omitted from now on.
Luckily, only few of the many terms in this equation are different from zero. Indeed, since
$\langle \delta\!\phi_a(\vec{r})\,\delta\!\phi_a(\vec{r}')\rangle = 
-\langle \delta\!\phi_a(\vec{r})\,\delta\!\phi_a(\vec{r}')\rangle$ and
$\langle \delta\!\phi_a(\vec{r})\,\delta\!\phi_c(\vec{r}')\rangle = 0$, we get $\langle \delta\!\phi_t(\vec{r})\,
\delta\!\phi_t(\vec{r}')\rangle = 0$ and $\langle \delta\!\phi_t(\vec{r})\,\delta\!\phi_j(\vec{r}')\rangle = 0$
where $j=a,c$. This follows directly from the equations (\ref{abcd_noise_a}) and (\ref{abcd_noise_c}). 
It is also clear that the noise term contribution $\zeta_{a_1}$ in Eq. (\ref{abcd_ave}) 
is exponentially suppressed and hence can be discarded. 
With this we end up with the following expression:
\begin{equation}
    \langle\delta\! a_1\rangle = - G\left[\frac{1}{4}\langle\delta\! \phi_a^2 - \delta\! \phi_c^2\rangle\right] = 
\frac{2}{\Phi_t}\dint{0}{t}{t_1}\int \, d^dx_1\,G_0^2(\vec{x}-\vec{x}_1,t-t_1)\langle f(a_i,c_i)(t_1)\rangle\,.
\end{equation}
from which we obtain that the approach to the equilibrium density of $A_1$ particles is again governed by
a power law, namely
\begin{equation}
    \langle\delta\! a_1\rangle = \frac{2}{\Phi_t}[a_1(\infty) - a_1(0)](8\pi D t)^{-d/2}\,.
\end{equation}
Following the same stepe as in Section II, we obtain for the autocorrelation function the expression
\begin{eqnarray}
    C(t,s) & = & \frac{(\Phi_t-\Phi_a)^2 - \Phi_c^2}{4\Phi_t^2}\langle \delta\! \phi_a(t)\delta\! \phi_a(s)
\rangle\nonumber\\ 
& = & \frac{(\Phi_t-\Phi_a)^2 - \Phi_c^2}{2\Phi_t^2}[a_1(\infty) - a_1(0)]\left[4\pi D(t+s)\right]^{-d/2}
\end{eqnarray}
with the same dependence on $s$ and $t$ as for the 
$A+A \rightleftharpoons C$ reaction scheme. Response functions are also calculated as previously, yielding again
the same functional dependences. For example, for the response of the $A_1$ particles to a perturbation that
conserves $\Phi_t$ we get
\begin{equation}
R_{\Phi_t}^{A_1}(t,s) =  \frac{2}{\Phi_t}[a_1(\infty) - a_1(0)]\left[8\pi D(t-s)\right]^{-d/2} \theta(t-s) ~.
\end{equation}

Let us finish this Section by very briefly discussing the general case $\mu\neq\lambda$. In principle, one
follows exactly the same steps as before, but the algebra is much more involved. Keeping only relevant terms 
we get instead of Eq. (\ref{abcd_a1}) the expression
\begin{eqnarray}
    \left(\der{}{t} -D\nabla^2\right)\, \delta\! a_1 & = & \sigma\, \delta\! a_1 - 
\frac{\sigma - \lambda\Phi_a}{2}\,\delta\!\phi_a + \frac{\mu \Phi_c}{2}\,\delta\!\phi_c\nonumber\\
    && - \frac{\mu}{4}(\delta\! \phi_a^2 - \delta\! \phi_c^2) - (\lambda -\mu)\delta\!a_1\,\delta\!\phi_a + 
(\lambda-\mu)\delta\!a_1^2 \label{abcd_a}~,
\end{eqnarray}
with $\sigma = \sqrt{\mu\lambda\Phi_t^2 + (\lambda-\mu)(\lambda\Phi_a^2-\mu\Phi_c^2)}$. This then yields the
following asymptotic approach toward the equilibrium density of the $A_1$ particles:
\begin{equation}
    \langle\delta\! a_1\rangle = g(\mu,\lambda,\Phi_a,\Phi_c) [a_1(\infty) - a_1(0)](8\pi D t)^{-d/2}\,,
\end{equation}
where we defined
\begin{equation}
    g(\mu,\lambda,\Phi_a,\Phi_c) = \frac{\lambda -3\mu}{4\sigma} + \frac{\lambda-\mu}{4\sigma^3}\left[(3\sigma-\lambda\Phi_a)^2 -(\mu\Phi_c)^2\right]\,.
\end{equation}
It is again to be noted that this leads to exactly the same functional forms for the autocorrelation and 
autoresponse functions.

\section{Conclusion}\label{conclusions}

In this work we extended the study of aging phenomena in reaction-diffusion systems
toward {\it reversible} reaction schemes. Our starting point thereby was the observation \cite{Zel77}
that in reversible reaction-diffusion systems slow dynamics, i.e., dynamics characterized
by power law relaxation, is generic. This is in stark contrast to the case of irreversible reactions as
here the approach to stationarity usually happens exponentially fast, with the exception of
nonequilibrium critical points which are governed by power laws. 

In our study we focused on simple models with reversible reactions whose behavior in the
asymptotic or scaling regime can be computed exactly \cite{Rey99}. We observed for all studied models
a simple scaling behavior of two-time response and correlation functions. Interestingly,
and in agreement with irreversible reaction-diffusion systems at  their critical point,
the two scaling exponents $a$ and $b$, see Equations (\ref{autoresp_scal}) and (\ref{autocorr_scal}),
were found to be different. The multi-species models we studied have the virtue that we can define
different responses, depending on the particle type for which an additional creation process is
considered and on the particle type whose density we are monitoring after the perturbation of the system.
We also studied a response where the perturbation conserves the total number of particles in the system.
For this perturbation we found that the exact result agrees with the expression one obtains from
the theory of local-scale invariance \cite{Hen01,Hen07a} that exploits the existence of space-time symmetries in aging
systems. However, if the system is perturbed in such a way that the total number of particles, a quantity
that is constant in the unperturbed system, is not conserved, than the response can not longer be cast in the
theoretically predicted form (\ref{lsi}). This result indicates that some care has to be taken
if one wants to apply space-time symmetries to cases where response functions result from perturbations
that change some otherwise conserved quantities.

It is very appealing that the theoretically predicted power law approach to stationarity is readily observed
in experimental systems \cite{Hup92,Sol01,Pin01}. 
It should therefore be possible to measure in these systems two-time quantities
in the scaling regime in order to verify the scenario of simple aging that follows from our study.

The models we consider in this work are to some extend artificial, as we allow multiple occupancy of a given 
site and only consider on-site reactions. It is an important question whether the scaling picture emerging
from the study of these simple models also holds in more realistic cases. Of special interest in this
context is the restriction to single-site reactions, as this disagrees with the actual experiments where
longer-range reactions prevail. Indeed, it has been stressed in the literature 
\cite{Rey99} that the models studied in this paper
yield as a stationary state a chemical equilibrium state. 
Voituriez et al. \cite{Voi05}, however, pointed out that distance-dependent
reversible reaction rates no longer yield asymptotically a chemical equilibrium state,
but that the stationary state is then a nonequilibrium state. This raises the interesting prospect that
one could be able to study the similarities and differences in the aging behavior of systems relaxing
toward equilibrium and nonequilibrium stationary states by changing the range of the reactions.
We plan to study this important aspect in our future work.

\appendix
\section{Relation between the asymptotic particle concentrations}

In the discussion of the aging processes for the $A + A\rightleftharpoons C$ reaction we extensively used the relation (\ref{steady}) between the equilibrium densities of the $A$ and $C$ particles. In this Appendix we briefly show that in zero dimension this condition may not be realized by all initial conditions.

Starting from the probability $P(m,n,t)$ for having at time $t$ $m$ resp. $n$ particles of type $A$ resp. $C$ at some lattice point, we can introduce the generating function (see, for example, \cite{Elg04,Elg06})
\begin{equation}\label{GF}
    \Psi(\tilde a,\tilde c,t)\equiv\sum_{m,n}\tilde a^m\tilde c^n P(m,n,t)\,.
\end{equation}
It then follows from the Master Equation (\ref{ME_singlepart}) that this generating function
is solution of the partial differential equation
\begin{equation}
    \der{}{t}\Psi = (\tilde a^2 - \tilde c)\left(\mu \der{}{\tilde c} -
    \lambda \der{^2}{\tilde a^2}\right)\Psi \,.
\end{equation}
It is more convenient to work with the shifted variables $\bar a\equiv \tilde a -1$ and $\bar c\equiv \tilde c -1$
which yields for the equation for the generating function the expression
\begin{equation}\label{quantum}
    \der{}{t}\Psi = (\bar a^2 + 2\bar a - \bar c)\left(\mu \der{}{\bar c} -
    \lambda \der{^2}{\bar a^2}\right)\Psi \,.
\end{equation}
The stationary solutions of this equation are given by
\begin{equation}\label{ground}
    \Psi_0(\bar a, \bar c) = \dint{{-\infty}}{\infty}{\bar z}\,
    G_0(\bar c, \bar a - \bar z)f(\bar z)\,,
\end{equation}
where $G_0$ is just the Green function for the one-dimensional diffusion equation,
\begin{equation}
    G_0(x, y) \equiv \left(4\pi \frac{\lambda}{\mu} x\right)^{-1/2}
    \exp\left(-\frac{y^2}{2\frac{\lambda}{\mu} x}\right)\,,
\end{equation}
and $f(\bar z)$ is any function satisfying the condition $f(0) = 1$ with only
\emph{positive} coefficients in the Taylor expansion. It directly follows that
there exist infinitely many solutions of the equation (\ref{quantum}), such as
\begin{equation}\label{non_poisson}
    \Psi_0(\bar a, \bar c) = 1 + \frac{\lambda}{\mu}\bar c +
    A \bar a + \frac{1}{2}\bar a^2\,,
\end{equation}
(with an arbitrary constant $A$) or
\begin{equation}\label{poisson}
    \Psi_p(\bar a, \bar c) = \exp\left(\frac{\lambda}{\mu}\bar c +
    \bar a\right)\,,
\end{equation}
which corresponds to the Poisson stationary distribution. 
It is now easy to see that the condition $\mu c_\infty = \lambda a_\infty ^2$ 
is indeed realized for the Poisson distribution (\ref{poisson}), but not for the distribution (\ref{non_poisson}).

\section{Perturbation series}

In their article \cite{Rey99} Rey and Cardy stress that one of the central points of their analysis
is the fact that the equation for the variable $c$ comes without explicit noise, thus implying that their
approach only works in cases where one of the equations of motion does not have an 
explicit noise dependence. In fact, this requirement is not really needed, as we demonstrate in this Appendix
by deriving the average number of $A$ particles directly from the `noisy' equation  (\ref{ALan}).

Indeed, starting from this equation we have
\begin{equation}\label{ALan2}
    (\partial_t - D_a\nabla^2 + \frac{1}{2}\sigma) \delta\!a = \frac{1}{2}\mu\delta\!\chi + \zeta - \lambda \delta\!a^2\,.
\end{equation}
The formal solution of the nonlinear equation
(\ref{ALan2}) is given by
\begin{equation}\label{A2}
    \delta\!a = G\left[\frac{1}{2}\mu\delta\!\chi + \zeta\right] - \lambda G\left[\delta\!a^2\right]\,,
\end{equation}
where the Green function $G$ is the inverse of the operator $\partial_t - D_a\nabla^2 + \frac{1}{2}\sigma$.
With this one can derive a perturbation series for, say, the average density $\langle\delta\!a\rangle$
following the same strategy as in [\onlinecite{Rey99}].
One thereby exploits the fact that the noise term contribution $G[\zeta]$ in this series is exponentially 
suppressed and hence can be discarded. To convince oneself that it is indeed so, one should consider terms 
such as $\langle G[\zeta]^2\rangle$ and $\langle G[\delta\!\chi] G[\zeta]\rangle$, which are exponentially 
small for late times $t$. Taking this into account, one gets
\begin{equation}\label{A_density}
    \langle\delta\!a\rangle = - \lambda \left(\frac{1}{2}\sigma\right)^{-3} \left(\frac{1}{2}\mu\right)^2\langle\delta\!\chi^2\rangle = -\frac{2\lambda\mu^2}{\sigma^3}\langle\delta\!\chi^2\rangle\,,
\end{equation}
which, of course, corresponds to the result
\begin{equation}
    \langle\delta\!a\rangle = - 2\langle\delta\!c\rangle\,.
\end{equation}
obtained in [\onlinecite{Rey99}] as the condition $\langle\delta\!\chi\rangle = 0$ holds.

\end{document}